\documentclass[aps,pra,amssymb,groupedaddress,showpacs]{revtex4}

\usepackage{graphicx}
\usepackage{dcolumn}
\usepackage{amsmath}
\usepackage{bm}
\begin{document}
\title{Quantum-classical correspondence of the Dirac matrices:\\ The
Dirac Lagrangian  as a Total Derivative}
\author{S. Savasta$^1$, O. Di Stefano$^{1\, , 2}$, and O.M. Marag\`{o}$^3$}
\affiliation{$^1$Dipartimento di Fisica della Materia e Ingegneria
Elettronica, Universit\`{a} di Messina Salita Sperone 31, I-98166
Messina, Italy.
\\ $^2$Dipartimento di Matematica, Universit\`{a} di Messina, Salita
Sperone 31, I-98166 Messina, Italy. \\
$^3$CNR-Istituto per i Processi Chimico-Fisici, Salita Sperone c.da
Papardo, I-98158 Faro Superiore Messina, Italy.}

\begin{abstract}
{The Dirac equation  provides a description of spin $1/2$ particles,
consistent with both the principles of quantum mechanics and of
special relativity. Often its presentation to students is based on mathematical propositions
that may hide  the physical meaning of its contents. Here we show that Dirac
spinors provide the quantum description of two unit classical vectors: one
whose components are the speed of an elementary particle and the
rate of change of its proper time
and a second vector which fixes the velocity direction. In this
context both the spin degree of freedom and antiparticles can be understood
from the rotation symmetry of these unit vectors. Within this approach the Dirac Lagrangian acquires a  direct physical meaning as the quantum operator describing the total time-derivative.
}
\end{abstract}
\maketitle

\section{Introduction}
Historically, Paul Dirac found the Klein-Gordon equation
physically unsatisfactory for the appearence of negative
probabilities arising from the second-order time derivative
\cite{Dirac}. Thus he sought for a relativistically invariant
Schr\"{o}dinger-like wave equation of first order in time derivative
of the form ($\hbar =c=1$):
\begin{equation}
    i\partial_t \psi =  H \psi\, .
\label{Sl}\end{equation} In order to have a more symmetric
relativistic wave equation in the 4-momentum components, he also
looked for an equation linear in space derivatives i.e. momentum
${\bf  P}= -i {\bm \nabla}$, so that $H$ takes the form
\cite{Dirac}:
\begin{equation}
 H =  {\bm \alpha} \cdot {\bf  P} +  \beta m \, , \label{DH}
\end{equation}
with ${\bm \alpha}$ and $\beta$ being independent on space, time and
4-momentum. The condition that Eq.\ (\ref{DH}) provides the correct
relativistic relationship involving energy, rest mass and momentum,
\begin{equation}
	E^2 = {\bf p}^2+m^2\, ,
\label{Ec}\end{equation}
requires that ${\bm \alpha}$ and $\beta \equiv \alpha_4$ obey the
anticommutation rules $\left \{ \alpha_i, \alpha_j \right \}= 2
\delta_{ij}$ ($i=1,4$) \cite{weinberg}. Dirac found that a set of $4
\times 4$ matrices satisfying this relation provides the lowest
order representation of the four $\alpha_i$. They can be expressed
as a tensor product of ($2 \times 2$) Pauli matrices $\rho_i$ and
$\sigma_i$ belonging to two different Hilbert spaces: ${\alpha}_i =
\rho_1 \sigma_i$ ($i=1,3$) and $\beta=\rho_3$ \cite{Dirac}.
As a consequence $\psi$ in Eq.\ (\ref{Sl}) is a 4-component wave
function. 
It turns out that ${\bm \alpha}$ is the velocity
operator\cite{Dirac,Breit}. So that $\psi^\dag {\bm \alpha} \psi$ is
the current density, determining the coupling with the
electromagnetic field \cite{Dirac,weinberg,peskin}

This derivation, although straightforward, does not help for a clear physical understanding of the four component wave function and of  Dirac matrices.
Indeed Richard Feynman in his Nobel lecture pointed out that {\em Dirac obtained his equation for the
description of the electron by an almost purely mathematical proposition. A
simple physical view by which all the contents of this equation can be seen is
still lacking}.

The fact that relativistic Dirac theory automatically includes spin
leads to the conclusion that spin is a purely quantum relativistic
effect originating from the finite dimensional representations of
the Lorentz group \cite{Fuchs}.  Nevertheless this
interpretation is not generally accepted, e.g. following Weinberg
argument (Ref.\ \cite{weinberg} Chapter 1) {\em \dots it is
difficult to agree that there is anything fundamentally wrong with
the relativistic equation for zero spin that forced the development
of the Dirac equation -- the problem simply is that the electron
happens to have spin $\hbar /2$, not zero.} Technically speaking,
the homogeneous Lorentz group (in contrast e.g. to the  group of
rotations) is not a compact group, hence the implementation of this
symmetry in quantum mechanics does not need the use of finite
dimensional representations. On the contrary, being non compact, it
has no faithful finite dimensional representation that is unitary
\cite{Fuchs,peskin}. Thus the homogeneous Lorentz group is the only
group of relativistic quantum field theory acting on
multiple-components quantum fields non-unitarily. This rather
surprising fact conflicts with a theorem proved by Wigner in 1931
(see Ref. \cite{weinberg} Chapter 2) which states that any symmetry
operation on quantum states must be induced by a unitary (or
anti-unitary) transformation.

The conflict is usually overcome,
either by regarding the field not as a multicomponent quantum
wavefunction but as a classical field \cite{peskin}, or by pointing
out that the fundamental group is not the (homogeneous) Lorentz
group but the Poincar\'e group \cite{weinberg}. Independently on the
point of view, one consequence is that the Hermitean conjugate
$\psi^\dag$ of the (four-component) spinor field $\psi$ does not
have the inverse transformation property of $\psi$ as requested by
quantum mechanics. The rather ad hoc, though generally accepted,
solution  is to define $\bar \psi = \psi^\dag \gamma^0$ called the
{\em Dirac conjugate} of $\psi$, being $\gamma^0\equiv \beta$ the
{\em time} Dirac matrix \cite{weinberg,peskin,maggiore,hey}.

In this paper we address some naturally
arising questions: is there a fundamental compact symmetry group
that requires the occurrence of spin? is there a reason why the
velocity operator ${\bm \alpha}$ of elementary matter particles is a vector made of $4
\times 4$ matrices instead of continuous variable operators? is
there a simple interpretation of Dirac four-component spinors? 
Here we discuss how
the spin of elementary particles can be understood as a consequence
of a rotation symmetry displayed by the kinematic variables
describing their space-time motion. Furthermore we show that such rotation
symmetry implies the existence of antiparticles.

Before starting our analysis, it is worth pointing out that here we follows the Dirac  {\em old fashioned} point of view, regarding the Dirac equation as a Schroedinger-like quantum mechanical equation providing the quantum description of a relativistic point-like particle. In contrast the modern view regards the Dirac equation as a classical wave equation describing the $1/2$ spin field. The field is then quantized by means of  canonical quantization for relativistic fields \cite{weinberg,peskin}.

\section{Rotations in quantum mechanics}

A rotation in classical physics is implemented by a $3 \times 3$
orthogonal matrix $R$ which, acting on a given vector ${\bf v}$,
gives the rotated vector ${\bf v}' = R {\bf v}$. Rotations are a
symmetry group whose generator is the angular momentum
\cite{Goldstein}. In quantum mechanics a symmetry transformation is
a transformation of the state kets describing the physical system
and of the operators $\hat O$ in the ket space. The transformation
operators ${\hat D}(R)$ are unitary and have the same group
properties as $R$ (see also supplementary information). The
expectation values of the angular momentum operators transform under
rotation as classical vectors: 
\begin{equation}
\left< J_k \right> \to R_{kl}\left<J_l \right>
\end{equation} 
where repeated indices are implicitly summed over \cite{Sakurai}. 
The lowest number $N=2j+1$ of dimensions in which
the angular momentum commutation relations are realized is $N=2$ (j
= 1/2). In this case the angular momentum operators can be
represented in terms of the Pauli matrices: $J_k = \sigma_k/2$.
Independently on the physical state in the 2D Hilbert space, they
obey the following relationship: 
\begin{equation}
\sum_i \langle \sigma_i \rangle^2=1
\end{equation}
which is not satisfied by higher order angular momentum
operators. Hence Pauli matrices are the {\em best} quantum
correspondents of classical unit vectors ${\bf \hat u}$ (with
$\sum_i {\hat u}^2_i = 1$) \cite{Sakurai}. If there is any classical
physical variable which is described by a unit vector, Pauli
matrices provide its natural quantization: ${\hat u}_j \to
\sigma_j$, which preserves rotation symmetry and ensures expectation
values which maps on the classical values.

\section{From relativistic kinematics to unit vectors}
Let us consider a relativistic point-like particle moving with
velocity $\mathbf{v}(t)$ with respect to an observer's reference
frame $K$ \cite{Jackson}. A possible spacetime worldline trajectory is shown in Fig.\ 1a.
We can consider the Lorentz-invariant quantity
\begin{equation}
d \tau(t)^2 \equiv dt^2 - d{\bf x}(t)^2\, .
\label{propertime}\end{equation}
In the instantaneous rest frame $K'$ of the particle $d{\bf x}=0$ and hence $d \tau(t)^2 = dt^2$ which admit solutions
$d \tau = \pm dt$.
The solution $d \tau = dt$ is the one usually presented in the introductory texts on special relativity (see e.g. \cite{Jackson}).
More recently Costella {\em et al.} \cite{Costella} pointed out that the negative solution provide a classical description of antiparticles
in agreement with the Feynman's interpretation of antiparticles as particles {\it going
back in  time} \cite{hey}.

Thus from Eq.\ (\ref{propertime}) it follows that:
\begin{equation}
d\tau = \pm \frac{d t}{\gamma(t)}\, ,
\end{equation}
where $\gamma=(1-v^2)^{-1/2}$ is the usual Lorentz boost parameter yielding time dilation.

The time $\tau$ is the proper time of the particle. Thus the
quantity $d\tau/d t \equiv \dot\tau= \pm 1/\gamma(t)$ expresses the rate
of change of the proper time with respect to the time $t$ of the
reference frame i.e. proper time velocity.

\begin{figure}[!ht]
\begin{center}
\resizebox{!}{6cm}{\includegraphics{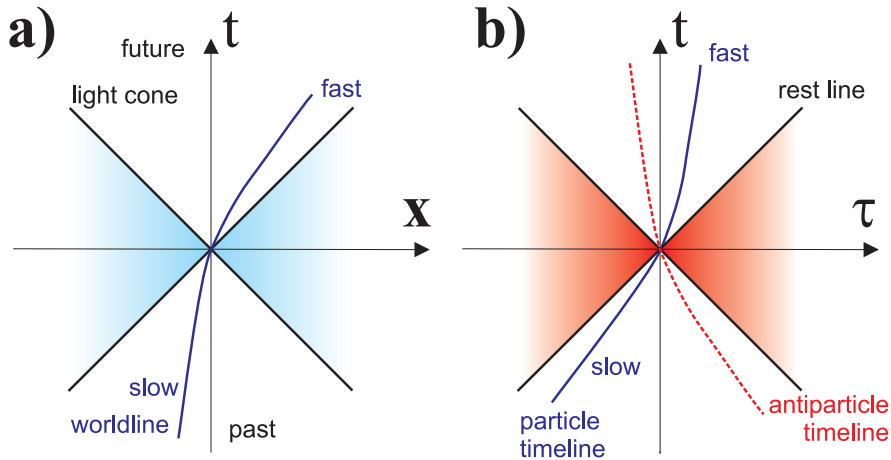}}
\caption{Worldlines of relativistic point-like particle and
antiparticles. a) In the usual Minkowski space-time diagram
particles and antiparticles worldlines are not distinguished. b) A
proper time-time diagram manifests the different kinematics of
particles ($\dot \tau >0$) and antiparticles ($\dot \tau <0$) with
respect to an observer's reference frame.}
\end{center}\label{worldlines}
\end{figure}

The usual way to describe spacetime trajectories (worldlines) relies on $t-{\bf
x}$ Minkowski diagrams, as shown in fig. \ref{worldlines}a, where
the light cones separates causal events from space-like events.
Additional information can be inferred by  drawing
proper time-time trajectories (timelines) as shown in fig. \ref{worldlines}b. The solid line describes the trajectory of a particle
with $\dot \tau >0$. Trajectories with
$\dot \tau >0$ can be identified by classical particle motion states, while
solutions with $\dot \tau <0$ correspond to classical anti-particle
states \cite{Costella}. 

When attempting to understand the spin and Dirac spinors on a
physical ground, the following question arises: {\em are there
kinematic variables which can be described as  components of unit
vectors?} 

\begin{figure}[!ht]
\begin{center}
\resizebox{!}{6cm}{\includegraphics{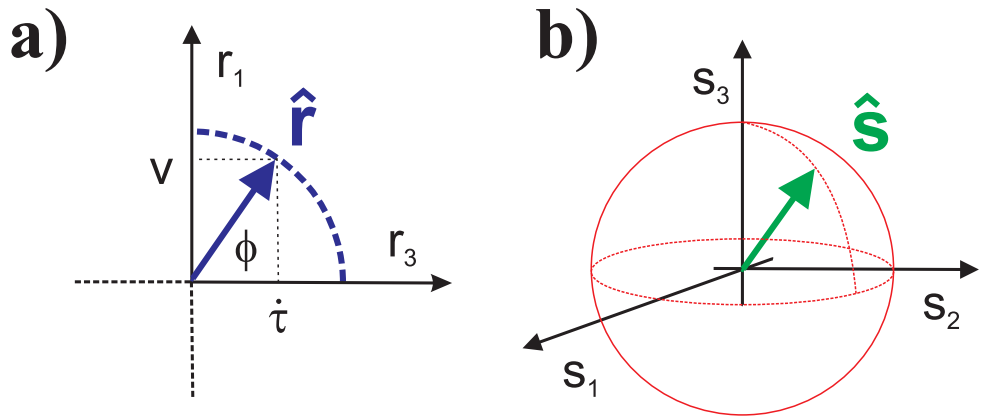}} \caption{Geometric
representations of unit vectors defining the kinematic state of a
relativistic point-like particle. (a) The unit vector ${\bf \hat r}$
is defined by the modulus of space and proper time velocities. A
particle changing its kinematic state will move on the unitary
circle defined by the angle $\phi$. (b) The unit vector ${\bf \hat
s}$ is defined by the spatial direction of velocity and can rotate
on the 3D unitary Poincar\'e sphere.}
\end{center}\label{unitvectors}
\end{figure}

The speed of a particle $v(t) = \left| {\bf \dot x}(t) \right|$ is
limited (by the speed of light $c =1$) like the component of a unit
vector. Additionally the quantity $\dot \tau$ can be viewed as the
complementary component of this unit vector. Indeed by definition of
$\gamma$, it follows that:
\begin{equation}
\left| {\bf \dot x} \right|^2 + \dot \tau^2 =1.\\
\end{equation}
Thus we can introduce a convenient unit vector ${\bf \hat r} \equiv(
\left| {\bf \dot x} \right| ,0,{\dot \tau})$ lying on the $r_1r_3$ plane (see Fig.
\ref{unitvectors}a) and an angle $\phi$ so that:
\begin{equation}
\left| {\bf \dot x} \right|=\sin \phi, \  \dot \tau =\cos \phi.\\
\end{equation}
A particle changing its kinematic state will move on the unitary
circle defined by the angle $\phi$.
Figure \ref{unitvectors}a displays one such kinematic vector with
positive components. If the particle is at rest with respect to the
reference frame, the vector lies on the $r_3$ axis (${\dot \tau}
= 1$). Moreover $\dot \tau$ decreases when the particle-speed increases
as predicted by special relativity (time dilation).

The unit vector ${\bf \hat r}$, besides ${\dot \tau}$, is able only
to describe the  modulus of the particle velocity $v = \left| {\bf
\dot x} \right|$. The particle velocity is indeed a 3D vector and
the direction of ${\bf \dot x}$ can be accounted for by one
additional 3D unit vector ${\bf \hat s}$ providing just the
direction of ${\bf \dot x}$. Hence the motion state of a particle
can  be described by a specific couple of unit vectors ${\bf \hat
r}$ and ${\bf \hat s}$:
\begin{equation}
{\dot \tau} = {r}_3 \, ,\hspace{1 cm} {\bf \dot x}= { r}_1\, {\bf
\hat s}\, .
\end{equation}

Within this approach changes of the particle velocity respect to an
inertial frame can be accounted for by rotations of ${\bf \hat r}$
in the kinematic plane  (changes of the  modulus) and rotations of
${\bf \hat s}$ (changes of the  direction). The unit vector ${\bf
\hat s}$ can be transformed according to arbitrary 3D rotations
around an arbitrary 3D unit vector ${\bf \hat n}$: 
\begin{equation}
 s_i \to  s_i' =
[{R}_{\bf \hat n}(\theta)]_{ij}\, s_j,
\end{equation}
where $\theta$ labels the angle of
rotation about ${\bf \hat n}$. On the other hand, physical
kinematic states ${\bf \hat r}$ admit only rotations about the
$r_2$-axis: 
\begin{equation}
r_i \to r_i' = [R'_{{\bf j}}(\phi)]_{ij}\, r_j\, .
\end{equation}

We observe that this rotation
symmetry suggests that states obtained rotating ${\bf \hat r}$, and
${\bf \hat s}$ should be considered as accessible states. In
particular ${\bf \hat r}$ also describes kinematic states in the
second and third quadrant with ${\dot \tau} < 0$, corresponding to classical antiparticles \cite{Costella}

From the point of view of classical special relativity, the
description in terms of ${\bf \hat s}$ and ${\bf \hat r}$ appears to
be redundant: a given velocity is described by two different states.
For example a particle with given velocity along direction ${\bf
\hat d}= {\bf \dot x}/ \left| {\bf \dot x} \right|$ can be described
by the unit vectors ${\bf \hat s}_\uparrow = {\bf \hat d}$ and ${\bf
\hat r}_\uparrow = (\sin \phi ,0,\cos \phi)$ with $\phi = \arcsin
\left| {\bf \dot x} \right|$, or equivalently by the unit vectors
${\bf s}_\downarrow =-{\bf \hat d}$ and ${\bf \hat r}_\downarrow =
(\sin \phi' ,0,\cos \phi')$ with $\phi' = -\phi$. This
two-valuedness can be described in terms of the helicity variable $h
= ({\bf \hat s} \cdot {\bf p})/p$. Below we show that this classical
twofold degeneracy is indeed the classical correspondent of the
helicity states determined by quantum spin. It is worth pointing out
that, although the present approach describes a spin-like degree of
freedom yet at a classical level, the interaction of a classical
particle with the electromagnetic field is not affected by this
additional degree of freedom, in contrast to what happens after
quantization.

\begin{figure}[!ht]
\begin{center}
\resizebox{!}{6 cm}{
\includegraphics{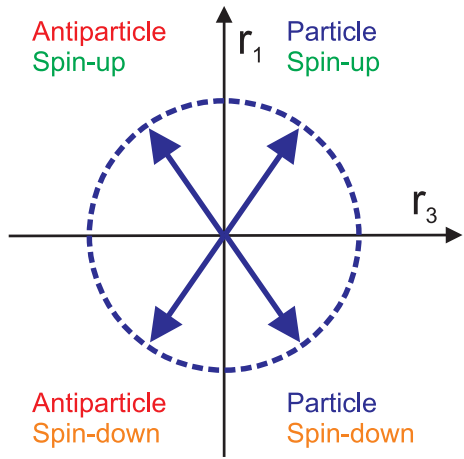}}
\caption{Sections of the {\em velocity space} describing the
different solutions. Unit vectors in the first and fourth quadrants
describe spin up and spin down particles respectively. Unit vectors
in the second and third quadrant describe spin up and spin down
antiparticles.} \label{tintegrated ky=0}
\end{center}
\end{figure}

Figure\ 3 provides a clear geometric interpretation of the different
kind of kinematic states: the first quadrant contains spin up
particles, the second one spin up antiparticles, the third spin down
antiparticles and the fourth quadrant spin down particles (see
Supplementary Information).

\section{From unit vectors to Dirac matrices}
As discussed earlier, a quantum mechanical description of unit
vectors is obtained replacing the classical vector components with
Pauli matrices: $r_i \to \rho_i$ and $s_i \to \sigma_i$, where
$\rho_i$ and $\sigma_i$ are now Pauli matrices acting on two
different Hilbert spaces $\mathcal{R}$ and $\mathcal{S}$. Hence:
\begin{eqnarray}
{\dot \tau} &=& {r}_3  \to \rho_3 \otimes I = \beta \nonumber\\
{\bf \dot x}&=&  {r}_1\, {\bf \hat s} \to \rho_1 \otimes {\bm
\sigma} = {\bm \alpha}\, , \label{corr1}\end{eqnarray} being $I$ the
identity operator in the $s$ space describing the direction of
velocity. The ${\bm \alpha}$ and $\beta$ are indeed the well-known
Dirac matrices in the standard representation \cite{Dirac}. Thus we
find a direct symmetry-driven correspondence showing that ${\bm
\alpha}$ is the velocity vector operator, and $\beta$ the
proper-time velocity operator. In this quantum picture the helicity
operator for a particle of given momentum becomes $h = ({\bf \sigma}
\cdot {\bf p})/p$ in agreement with the Dirac theory \cite{peskin}
and with our classical derivation of helicity states. In summary, at
a classical level, the rotation symmetries we introduced describe
the correct relativistic kinematics. Furthermore they lead to two
possible helicities for a given motion state and the existence of
particles with $\dot \tau < 0$. After quantization, without any
other additional assumption, they give rise to the 4-component Dirac
spinors fully describing quantum spin and antiparticles with the
corresponding expectation value $\langle \beta \rangle < 0$. We
point out that in this framework the Dirac matrices are derived only
on the basis of the fundamental rotation symmetry without any
reference to the Dirac equation or to the energy of a relativistic
free particle.

At this point it is interesting to show that the energy of a relativistic free particle can be written as observed  By Breit as $E = {\bf v} \cdot {\bf p} + \dot \tau m$. This equation has the same structure of the Dirac Hamiltonian.

\section{The Dirac Lagrangian as a Total Derivative}
A crucial feature of this symmetyry based derivation of the Dirac matrices
is that it has been obtained by
regarding the position-coordinate and the proper time of the
particle on the same footing as functions of the time-coordinate of
the reference frame, i.e. we use $t$ as the relevant {\it meter} to
which compare the dynamical evolution of all other observables (as
happens in non-relativistic mechanics). Accordingly if a quantum
wavefunction depends on position ${\bf x}$,  it should
also depend on $\tau$ i.e. $\psi(t,{\bf x},\tau)$. We now consider a
particle with such wavefunction and require the conservation of its
probability density $\mathcal{D}(t,{\bf x},\tau) \equiv
\psi^*(t,{\bf x},\tau) \psi(t,{\bf x},\tau)$:
\begin{equation} \frac{d }{dt}\,  \mathcal{D} =
\frac{d \psi^*}{dt} \psi + \psi^*  \frac{d \psi}{dt} = 0\, .
\label{Liouville}\end{equation} Stationarity of $\psi$ (and
$\psi^*$) ensures that a continuity equation holds. In
turn if, as in Lagrangian formulations with complex
fields, $\psi$ and $\psi^*$ are regarded as independent,
Eq.\ (\ref{Liouville}) implies stationarity of $\psi$ (and of
$\psi^*$) i.e. $d \psi/dt=0$. Because the time dependence of the
wavefunction is both explicit and implicit (through the time
dependence of ${\bf x}$ and $\tau$), we have:
\begin{equation}
   \frac{d}{dt} \psi = \partial_t\, \psi  +
    {\bf \dot x} \cdot \partial_{\bf x}\, \psi
    + {\dot \tau} \partial_\tau\, \psi\, =0.
\end{equation}
Thus the request of stationarity and the crucial symmetry-based
quantization Eq.\ (\ref{corr1}), provides a generalized Dirac-like
equation:
\begin{equation}
    i {\partial}_t \psi(t,{\bf x},\tau) = H_g \psi(t,{\bf x}, \tau)\, ,
\label{gDe}\end{equation}
with
\begin{equation}
H_g = {\bm \alpha} \cdot (-i \partial_{\bf x}) + \beta (-i
\partial_\tau)\, . \label{Hg}\end{equation}
The corresponding Lagrangian density can be written as
\begin{equation}
	\psi^\dag {\cal L} \psi = \psi^\dag i(\partial_t+ {\bm \alpha} \cdot {\partial}_{\bf x} +  \beta  {\partial}_\tau) \psi \, .
\label{gLag}\end{equation}
This Lagrangian operator ${\cal L}$ has a precise physical meaning, being the Hermitean quantum operator describing the total time derivative:  $i d / dt \to {\cal L}$ after symmetry-based
quantization Eq.\ (\ref{corr1}).
Equation (\ref{Hg}) is more
general than the Dirac equation since the mass parameter is now replaced
by the internal time momentum operator. Elementary matter particles
of given mass can  be viewed as eigenstates of this operator. In
this case Eq.(\ref{Hg}) reduces to the standard Dirac equation. On
the other hand, this new degree of freedom offers a chance to
shed new light on some fundamental aspects and concepts.

As an example this new degree of freedom could be exploited to discuss charge conservation.
Charge conservation is generally accounted by invariance
under phase change of a field, Eq.\ (\ref{Hg}) allows the interpretation of this phase
change as the consequence of proper time translation applied to an
eigenstate of the proper time operator. Charge conservation could thus be viewed as arising from invariance of the Lagrangian under proper time translations.

It is also interesting to address the proper time reversal
symmetry $\tau \rightarrow -\tau$. Applying this symmetry to a
positive energy solution of Eq. \ref{Hg} which is also an eigenstate
of the proper time momentum (i.e. a particle with fixed mass and
charge), we obtain its antiparticle with positive energy and
opposite charge, so  antiparticle solutions with positive energy emerge without the need for second
quantization adjustments. 
\section{Conclusions}
Within the approach here presented, changes of the velocity modulus and direction of a relativistic particle can
simply be accounted by rotations of two independent unit vectors. Dirac spinors just provide the quantum description
of these rotations. These transformations are able to describe the helicity of a pointlike particle yet at a classical
relativistic level, rendering spin a less mysterious degree of freedom. This analysis sacrifies explicit covariance by
making explicit rotation symmetry which nevertheless is the fundamental symmetry on which the algebra of the
Lorentz group is based.
We observe that the addressed rotation symmetries form a compact group hence with unitary finite-dimensional
representations as all other symmetry groups in quantum field theory. A feature of this analysis is that it has been
obtained by regarding the position-coordinate and the proper time of the particle on the same footing as functions
of the time-coordinate of the reference frame, i.e. t is used as the relevant meter to which compare the dynamical
evolution of all other observables (as happens in non-relativistic quantum mechanics).

Within this approach we derived the Dirac equation just invoking
total stationarity of the wavefunction with respect to the reference
time. No assumptions about the classical relativistic energy of a
particle or about the quantum operator replacements ($E \to i
\partial_t$ and ${\bf p} \to -i \partial_{\bf x}$) have been
performed. 
Finally we observe that the quantum replacement $m \to -i
\partial_\tau$ implies an internal time-energy uncertainty
principle $\Delta \tau\, \Delta m$, e.g. as required by a gedanken
experiment recently proposed by Aharonov and Rezni \cite{Ahranov}.



\end{document}